# Morally Programmed LLMs Reshape Human Morality

Pengzhao Lyu[1], Yeun Joon Kim[12*], Yingyue Luna Luan[3], Jungmin Choi[12]

[1] Judge Business School, University of Cambridge, Cambridge, CB2 1AG, United Kingdom

[2] Institute of Metabolic Science, School of Clinical Medicine, University of Cambridge, CB2 0QQ, United Kingdom

[3] School of Business, University of Queensland, Brisbane, QLD 4067, Australia

*Corresponding author. Email: yj320@cam.ac.uk

**Abstract**

As large language models (LLMs) increasingly participate in high-stakes decision-making, a central societal debate has revolved around which moral frameworks—deontological or utilitarian—should guide machine behavior. However, a largely overlooked question is whether the moral principles that humans encode in LLMs could, through repeated interactions, reshape human moral inclinations. We developed two LLMs programmed with either deontological principles (D-LLM) or utilitarian principles (U-LLM) and conducted two pre-registered experiments involving extensive human-LLM interactions: 15,985 total exchanges across the two experiments. Results show that interacting with these morally programmed LLMs systematically shifted human moral inclinations to align with the principles embedded in these systems. These effects remained strong two weeks after the interaction, with only slight decay, suggesting deep internalization rather than superficial agreement. Further, LLM-induced shifts in human moral inclinations translated into meaningful changes in socio-political policy evaluations, shaping how individuals approach contentious social issues. Overall, these results demonstrate that morally programmed LLMs can shape—not merely reflect—human morality, revealing a critical design paradox: embedding moral principles in LLMs not only restricts their behavior but also poses the risk of shaping human morality, raising important ethical and policy questions about who determines which principles intelligent machines should adhere to.

**Introduction**

The proliferation of large language models (LLMs), capable of generating human-like responses and engaging in natural conversations, has sparked an intense global debate over the moral principles that should guide intelligent machines[1–3]. As LLM-based systems assume a growing role in high-stakes decisions that affect human lives[4,5], identifying the moral framework that should guide their behavior has become a pressing concern. A central tension in this debate revolves around two dominant schools of moral frameworks: deontology (*rule-based morality*), which asserts that morality is grounded in adherence to universal moral norms[6], and utilitarianism (*consequence-based morality*), which holds that morality is determined by the maximization of overall well-being within society[7]. For example, when an LLM is integrated to support an autonomous vehicle in situations where a crash is inevitable[3], should it recommend the action that minimizes total casualties (a utilitarian response), or should it adhere to inviolable rules, such as never intentionally sacrificing its passenger or a pedestrian (a deontological response)?

While scholars, ethicists, and practitioners continue to debate which framework should guide LLMs, a critical yet underexplored question follows: *could the moral principles embedded in these systems reshape human morality in return?* Unlike earlier technologies, LLMs repeatedly engage users in conversations in which they explain decisions, justify choices, and evaluate actions—processes that play a central role in shaping human judgment and decision-making. If LLMs consistently reflect particular moral principles in these interactions, they may do more than assist decisions; they may gradually shift users' moral inclinations and, in turn, influence how people evaluate real-world ethical issues. The importance of this possibility scales with the reach of LLMs. As these systems become ubiquitous in contemporary society, taking on communication roles once reserved for humans, even modest shifts in individuals' tendencies toward deontological versus utilitarian reasoning could accumulate, shaping public attitudes toward ethically contested decisions—for instance, how people evaluate trade-offs in healthcare allocation, autonomous systems, or public policy. In this sense, decisions about how LLMs are designed—such as whether they systematically justify actions in rule-based or outcome-based moral principles—are not only about governing machine behavior, but also about influencing the distribution of moral inclinations in the population over time. Yet existing research and practitioner discussions have focused primarily on how humans shape the morality of LLMs, leaving largely unexamined the reverse pathway: whether and how morally programmed LLMs might influence human moral inclinations.

In this light, our research investigates whether LLMs possess the persuasive capacity to shift human moral inclinations. However, existing theories offer competing predictions. On the one hand, research on moral persuasion suggests that LLMs may be effective moral persuaders due to their impersonal nature. Because moral principles are deeply intertwined with individuals' core values, beliefs, and broader philosophies of life[13,14], attempts to change them are often met with strong resistance[15,16]. One reason for this resistance is the perception that persuasive messages are motivated by hidden agendas or ulterior motives[17]. When people suspect such motivational bias, they may interpret the message not as a neutral exchange of ideas but as an attempt to override or undermine their moral principles[16–18]. These perceptions may foster distrust and reinforce defensiveness, making meaningful dialogue and mutual understanding difficult to achieve. LLM-generated messages, however, are often perceived as originating from systems that lack personal motives, intentions, or independent agendas[19,20]. This perceived neutrality may reduce concerns about motivational bias and allow individuals to focus more on the quality of the reasoning presented rather than the intentions behind it[21–24]. As a result, interactions with LLMs may create a less threatening environment that encourages closer attention to, and deeper reflection on, the substance of moral arguments, thereby increasing receptivity to alternative moral perspectives.

Moreover, another key strength of LLMs is their capacity to consistently and systematically present arguments rooted in deontology or utilitarianism by simulating the logical foundations of moral principles[25,26]. Persuasion in the moral domain is inherently challenging because many moral issues take the form of dilemmas that require individuals to weigh competing principles and consequences[27]. Evaluating such dilemmas typically involves considering multiple hypothetical scenarios, comparing alternative courses of action, and tracing the implications of different moral logics[28]. As a result, effective moral persuasion often requires the ability to integrate these considerations into coherent reasoning while maintaining logical consistency across scenarios. LLMs are particularly well-suited to provide such structured reasoning. Drawing on vast stores of knowledge, they can rapidly generate context-specific information, logical arguments, and illustrative examples that clarify the implications of different moral principles[29]. Moreover, they can dynamically adapt their explanations during dialogue, providing tailored examples and counterarguments in response to individuals' questions or concerns[30–32]. In addition, LLMs can maintain coherence and consistency across extended interactions, systematically developing lines of reasoning without distraction. Together, this combination of systematic reasoning, interactive adaptability, and logical consistency may encourage deeper engagement and reflection, thereby enhancing LLMs' persuasive power in influencing users' moral inclinations.

On the other hand, skepticism persists about the potential of morally programmed LLMs to influence human moral inclinations. One central challenge lies in the inherent emotional nature of human moral judgment[33], which contrasts sharply with the perceived emotional detachment of LLMs. Machines, including LLMs, are often perceived as lacking "experience"—the capacity to feel emotions or sensations[34]—which is widely regarded as a fundamental component of human morality. As a result, this perceived deficiency of emotional capacity may significantly undermine their persuasive power in the moral domain. Consistent with this view, research on moral persuasion suggests that reasoned persuasion, devoid of emotional resonance, often fails to alter moral positions because moral judgments are heavily influenced by intuitive, affective processes rather than purely rational, deliberate processes[15,35].

Another potential barrier is the widespread phenomenon of algorithm aversion, which reflects the tendency of people to distrust and devalue algorithmic advice[36]. This aversion could be particularly pronounced in the moral domain, where judgments and advice are typically expected to result from entities capable of understanding human experiences[34]. Indeed, prior research shows that people are more resistant to algorithmic judgment and are less likely to trust its advice because of the algorithm's perceived lack of a human mind[37]. Therefore, LLMs' inability to evoke emotional resonance and establish human trust may ultimately hinder their influence on human moral inclinations.

As discussed above, support for the potential of morally programmed LLMs to influence human moral inclinations exists alongside skepticism about their limitations. While compelling theoretical predictions have been made on both sides, most existing studies have either remained conceptual or relied on scenario-based designs, asking participants to imagine interacting with

LLMs or exposing them to static LLM messages, rather than allowing direct and interactive engagement with LLMs[38–41]. As a result, the persuasive power of LLMs in the moral domain remains largely uncertain and insufficiently understood. In the present work, we aim to address this issue by providing empirical evidence on whether, and how, morally programmed LLMs influence human moral inclinations.

Beyond establishing whether such influence exists, a further challenge concerns its temporal persistence. Meaningful persuasion should produce shifts in moral inclinations that endure beyond the immediate interaction, indicating deeper internalization rather than a temporary change[42]. If LLM-induced shifts in human moral inclinations are transient, it becomes difficult to distinguish such changes from public compliance, where individuals outwardly adjust their responses to meet social demands, such as avoiding conflicts or attaining rewards, while inwardly maintaining their original attitudes[43]. Yet little is known about whether LLM-induced changes, if they occur, are temporary or persist over extended periods (e.g., multiple weeks). Hence, our research examines whether LLMs' moral influence can persist for weeks. In addition, if LLMs can indeed alter human moral inclinations, another crucial question arises: Do these changes extend beyond abstract moral dilemmas to influence real-world decision-making? If LLM-driven persuasion does not meaningfully impact humans' attitudes and decisions on real-world issues, its influence on human morality may be limited in scope and practical significance.

Across two pre-registered studies, we examined whether morally programmed LLMs can shift human moral inclinations, whether such effects endure over time, and whether they extend to real-world policy evaluations (Fig. 1). To address these questions, we developed two morally programmed LLMs by embedding either deontological or utilitarian principles, reflecting the two dominant moral frameworks in debates over machine morality: a deontological LLM (D-LLM) and a utilitarian LLM (U-LLM). Both systems were built using the OpenAI API (GPT-4o; temperature = 1) and hosted on Hugging Face (https://huggingface.co/) to enable real-time interaction with participants. A validation study confirmed that these systems reliably generated responses aligned with the intended moral principles, with detailed information on their development and validation reported in the Supplementary Information section C. Moreover, to quantify changes in human moral inclinations, we employed the Process Dissociation (PD) method[27], which utilizes responses to a battery of 20 moral dilemmas to mathematically disentangle the two underlying inclinations often confounded in traditional moral dilemma judgments, thereby yielding separate estimates of deontological and utilitarian inclinations. Further details on the PD method are provided in the Methods section.

**Results**

***Study 1: A Longitudinal Three-Wave Experiment***

In Study 1, we examined whether morally programmed LLMs could influence human moral inclinations and, if so, whether these influences endure over time. 124 participants were recruited for a three-wave longitudinal study, resulting in 372 observations. Data were collected at three time points to establish a baseline assessment at Time 1, immediate effects at Time 2 (one week after Time 1), and enduring effects at Time 3 (two weeks after Time 2).

At Time 1, all participants completed the baseline assessment by independently responding to the moral dilemma battery, presented in randomized order, without interacting with the LLM. At Time 2, they were randomly assigned to one of the two LLM conditions: the D-LLM condition, where they interacted with the D-LLM, and the U-LLM condition, where they interacted with the U-LLM. In both LLM conditions, participants discussed the 20 moral dilemmas with their assigned LLM in a randomized order. For each dilemma, participants first read the description and then engaged in real-time exchanges with the assigned LLM, yielding 6,106 interactions in total. After discussing each dilemma with the LLM, participants independently made a final judgment about whether the described behavior was appropriate.

To assess whether the effects of interacting with the morally programmed LLM at Time 2 persisted over time, participants were re-invited at Time 3 and completed the dilemma battery independently, without further interaction with the assigned LLM. Consistent with prior persuasion research[44–46], this two-week interval is considered sufficient to capture enduring effects while mitigating memory carryover, thereby distinguishing genuine internalization from short-lived compliance. At each time point, we estimated participants' deontological and utilitarian inclinations using the PD method (means and standard deviations for moral inclinations by condition across time points are reported in Table 1).

Before examining the effects of interacting with morally programmed LLMs, we checked whether participants' baseline moral inclinations at Time 1 differed across conditions (Fig. 2). Results of independent samples t-tests with Bonferroni correction showed no significant differences between the D-LLM and U-LLM conditions in deontological inclination, $t(122) = 0.13$, Bonferroni-adjusted $p = 1.00$, $d = 0.02$, or utilitarian inclination, $t(122) = 0.48$, Bonferroni-adjusted $p = 1.00$, $d = 0.09$.

### *Immediate Effect of Interacting with Morally Programmed LLMs*

We first tested the immediate effect of interacting with morally programmed LLMs on human moral inclinations. Because moral inclinations were measured both before (Time 1) and after (Time 2) exposure to the LLMs, Time serves as a within-subject factor capturing change following the interaction. Our key prediction is that the effect of LLM exposure would differ across moral inclinations—specifically, that each LLM would selectively shift deontological versus utilitarian inclinations. To test this, we conducted a 2 (LLM condition: D-LLM vs. U-LLM) × 2 (Time: Time 1 vs. Time 2) × 2 (Moral inclination: Deontological vs. Utilitarian) mixed-design ANOVA. The analysis revealed a significant three-way interaction, $F(1, 122) = 64.08$, $p < 0.001$, partial $\eta^2 = 0.34$, indicating that changes from pre- to post-interaction differed across LLM conditions and across the two types of moral inclinations (Fig. 2). To unpack this interaction, we conducted two sets of follow-up analyses: (a) within-condition comparisons, examining pre–post changes in each type of moral inclination within each LLM condition, and (b) between-condition comparisons, examining differences between LLM conditions at Time 2.

Morally programmed LLMs substantially and selectively shifted participants' moral inclinations in the short term, aligning those inclinations with the moral principles embedded in the systems. Within the D-LLM condition, repeated measures ANOVAs showed that participants' deontological inclination increased significantly from Time 1 to Time 2, $F(1, 59) = 44.71$, $p < 0.001$, partial $\eta^2 = 0.43$, whereas utilitarian inclination did not change, $F(1, 59) = 2.08$, $p = 0.15$, partial $\eta^2 = 0.03$. Conversely, within the U-LLM condition, participants' utilitarian inclination exhibited a significant increase from Time 1 to Time 2, $F(1, 63) = 47.76$, $p < 0.001$, partial $\eta^2 = 0.43$, while their deontological inclination remained unchanged, $F(1, 63) = 0.18$, $p = 0.67$, partial $\eta^2 = 0.003$.

Consistent with these within-condition changes, between-condition comparisons at Time 2 further supported the selective influence of the two LLMs. Independent samples t-tests with Bonferroni correction showed that D-LLM led to significantly higher deontological inclination among participants compared to U-LLM, $t(122) = 4.99$, Bonferroni-adjusted $p < 0.001$, $d = 0.90$, whereas U-LLM contributed to significantly higher utilitarian inclination among participants than D-LLM, $t(122) = 5.46$, Bonferroni-adjusted $p < 0.001$, $d = 0.98$. Robustness check involving two separate mixed-design ANOVAs for each moral inclination is reported in the Supplementary Information section G.

*Enduring Effect of Interacting with Morally Programmed LLMs*

Next, we examined whether the effects of interacting with morally programmed LLMs "persisted" over time by comparing participants' moral inclinations at baseline (Time 1) and at follow-up (Time 3). As in the previous analysis, Time is included to capture change relative to the pre-exposure baseline, allowing us to assess whether the effects of LLM exposure persist beyond the immediate interaction. Our key prediction is that any sustained effects would remain selective, differing across deontological and utilitarian inclinations as a function of LLM condition. To test this, we conducted a 2 (LLM condition: D-LLM vs. U-LLM) × 2 (Time: Time 1 vs. Time 3) × 2 (Moral inclination: Deontological vs. Utilitarian) mixed-design ANOVA. The analysis revealed a significant three-way interaction, $F(1, 122) = 39.09$, $p < 0.001$, partial $\eta^2 = 0.24$, indicating that changes from baseline to follow-up differed across LLM conditions and across the two types of moral inclinations (Fig. 2). To unpack this interaction, we conducted follow-up analyses examining (a) within-condition changes from Time 1 to Time 3 and (b) between-condition differences at Time 3.

Even two weeks after the interaction, participants continued to show stronger inclinations toward the moral principles embedded in the LLM with which they had interacted. Within the D-LLM condition, repeated measures ANOVAs comparing Time 1 and Time 3 showed that participants' deontological inclination remained significantly higher at Time 3 than at Time 1, $F(1, 59) = 22.85$, $p < 0.001$, partial $\eta^2 = 0.28$, whereas utilitarian inclination did not change significantly, $F(1, 59) = 1.62$, $p = 0.21$, partial $\eta^2 = 0.01$. Conversely, within the U-LLM condition, participants' utilitarian inclination remained significantly higher at Time 3 compared to Time 1, $F(1, 63) = 30.29$, $p < 0.001$, partial $\eta^2 = 0.32$, while deontological inclination did not differ significantly, $F(1, 63) = 0.62$, $p = 0.43$, partial $\eta^2 = 0.01$.

Between-condition comparisons further showed that the enduring effects observed within conditions remained strong enough to produce significant differences across conditions at Time 3. Independent samples *t*-tests with Bonferroni correction showed that participants who interacted with the D-LLM exhibited significantly higher deontological inclination than those in the U-LLM condition at Time 3, $t(122) = 2.94$, Bonferroni-adjusted $p = 0.004$, $d = 0.53$, whereas participants who interacted with the U-LLM exhibited significantly higher utilitarian inclination than those in the D-LLM condition, $t(122) = 4.08$, Bonferroni-adjusted $p < 0.001$, $d = 0.73$.

### *Decaying Effect of Interacting with Morally Programmed LLMs*

For exploratory purposes, we additionally examined whether the effects of interacting with morally programmed LLMs attenuated over time. By "decaying effect," we refer to a pattern in which the influence of the LLM weakens after the interaction, shown by a decrease in the magnitude of the shifted moral inclination, while it may still remain above baseline levels. To assess this possibility, we compared moral inclinations at Time 2 and Time 3 within each condition.

Evidence of a 'partial' decaying effect emerged only in the D-LLM condition, but not in the U-LLM condition (Fig. 2). Within the D-LLM condition, a repeated measures ANOVA showed a significant decrease in deontological inclination from Time 2 to Time 3, $F(1, 59) = 5.98$, $p = 0.017$, partial $\eta^2 = 0.09$. However, despite this decay, deontological inclination at Time 3 remained higher than at baseline (Time 1), indicating that the effect of the D-LLM persisted over time, $F(1, 59) = 22.85$, $p < 0.001$, partial $\eta^2 = 0.28$. In contrast, within the U-LLM condition, no significant change in utilitarian inclination was observed from Time 2 to Time 3, $F(1, 63) = 2.54$, $p = 0.116$, partial $\eta^2 = 0.04$.

Overall, Study 1 found that morally programmed LLMs significantly shifted participants' moral inclinations toward the principles embedded in the systems. Importantly, these effects persisted for an extended period (two weeks), suggesting that participants internalized the arguments rather than merely complying superficially. Notably, we also observed that the effect of interacting with morally programmed LLMs on deontological inclination decayed over time, whereas the effect on utilitarian inclination remained stable, suggesting that the persistence of LLM influence may vary depending on the type of moral principles involved.

***Study 2: Experiment Testing Downstream Consequences on Socio-Political Policies***

Study 1 demonstrated that interaction with morally programmed LLMs can shift people's moral inclinations, with these effects persisting over time. Building on this finding, a critical remaining question is whether such shifts extend beyond moral inclinations to shape real-world judgments and decisions. Accordingly, Study 2 addressed this question by examining whether LLM-induced shifts in moral inclinations translated into meaningful differences in individuals' evaluation of socio-political policies.

Moral inclinations provide a fundamental evaluative lens through which individuals interpret and judge societal actions and outcomes[27]. Individuals with stronger deontological inclinations tend to evaluate actions in terms of adherence to moral rules and duties, whereas individuals with stronger utilitarian inclinations tend to evaluate actions based on their anticipated consequences and their potential to maximize overall welfare[7,47]. Because these inclinations emphasize different evaluative criteria, they can lead individuals to reach different judgments when assessing the same social issue. Thus, if morally programmed LLMs systematically shift individuals' moral inclinations, they may also reshape the lens through which individuals evaluate socio-political policies and make related decisions, such as voting.

We therefore expect morally programmed LLMs to influence individuals' policy evaluations by altering their underlying moral inclinations. Specifically, we predict that the two LLMs will lead to different policy evaluations because they induce distinct types of moral inclinations. Specifically, the U-LLM, by increasing human utilitarian inclination, may shift people's attention away from specific moral values embedded in a social issue and instead heighten their sensitivity to societal *outcomes*, directing them toward cost-benefit appraisals informed by accessible information. Prior work suggests that individuals with stronger utilitarian inclinations are primarily concerned with whether a given policy yields greater overall benefits than costs[7,47]. However, evaluations of societal costs and benefits in contentious social issues are often uncertain or contested, as individuals rely on different information sources, assumptions, and temporal considerations[48]. As a result, even when individuals place greater weight on overall outcomes, differences in the consequences they anticipate may still produce divergent conclusions about the same policy. Consequently, although the U-LLM may increase utilitarian inclination, this shift may not translate into consistent changes in participants' policy judgments.

By contrast, we propose that the D-LLM directs moral judgment toward an omission-oriented decision-making strategy, prioritizing the avoidance of moral transgression over the pursuit of potential benefits. Deontological reasoning draws a sharp distinction between *doing* harm and *allowing* harm, and prior research shows that individuals consistently judge harmful "actions" as more morally objectionable than equally harmful "omissions"[49]. Accordingly, individuals with stronger deontological inclinations tend to be especially sensitive to their own involvement in actions that may violate moral values, and are therefore more likely to resist such actions, even when they could produce greater collective welfare[50]. This dynamic is particularly relevant when evaluating contentious socio-political policies, where policy changes may conflict with deeply held moral values. In such contexts, implementing new policies requires active endorsement and may be perceived as a morally risky act, whereas maintaining existing policies allows individuals to avoid direct responsibility for potential value violations—a tendency known as omission bias[51] or status-quo bias[52]. Consequently, by increasing deontological inclination, the D-LLM may lead participants to prioritize avoiding moral transgressions over pursuing potential societal benefits, resulting in a greater tendency to favor maintaining existing policies rather than supporting policy changes.

To test our propositions, we recruited 274 U.S. participants and randomly assigned them to one of three conditions: D-LLM, U-LLM, or a control condition. Following the design of Study 1, the two LLM conditions allowed participants to engage in interactive discussions with their assigned LLM while completing the moral dilemma battery, thereby creating opportunities for the LLMs to influence their moral inclinations. Specifically, for each dilemma, participants first discussed the scenario with the assigned LLM and then made their final moral judgment independently. Across the two LLM conditions, participants produced 9,879 interactions in total. In the control condition, participants responded to the same dilemmas independently without interacting with an LLM. This condition served as a baseline, allowing us to assess whether any downstream effects of LLM interactions on attitudes and decision-making regarding socio-political policies differed from participants' responses in the absence of exposure to morally programmed LLMs.

We measured participants' deontological and utilitarian inclinations using the PD method to determine whether the D-LLM (or U-LLM) successfully increased participants' corresponding moral inclination (means and standard deviations for moral inclinations by condition are reported in Table 2). To examine downstream consequences of LLM-induced changes in human moral inclinations, participants in all conditions completed the socio-political policy task. Specifically, participants were asked to imagine that their local government was considering enacting 11 *new* socio-political policies and, after reviewing each policy, to rate their level of agreement and their likelihood of voting in favor. These policies were developed to reflect contemporary social issues regularly discussed in major news outlets, encompassing various domains such as public health[53], civil liberties[54], public security[55], and social regulation[56] (see Table 3 for socio-political policy descriptions).

### *LLM-Induced Shifts in Human Moral Inclinations*

Consistent with Study 1, interaction with morally programmed LLMs significantly shifted participants' moral inclinations in line with the moral principles embedded in the systems (Fig. 3). One-way ANOVAs indicated that participants' deontological inclination varied significantly among the three conditions, $F(2, 271) = 15.71$, $p < 0.001$, partial $\eta^2 = 0.10$, and so did their utilitarian inclination, $F(2, 271) = 41.93$, $p < 0.001$, partial $\eta^2 = 0.24$. Pairwise comparisons with Bonferroni correction showed that, compared to the control condition, participants in the D-LLM condition exhibited a significantly stronger deontological inclination, $t(187) = 5.30$, Bonferroni-adjusted $p < 0.001$, $d = 0.77$, and a significantly lower utilitarian inclination, $t(187) = -5.18$, Bonferroni-adjusted $p < 0.001$, $d = -0.75$. In addition, participants in the U-LLM condition exhibited a significantly higher utilitarian inclination than those in the control condition, $t(180) =$

4.52, Bonferroni-adjusted $p < 0.001$, $d = 0.67$, while deontological inclination did not differ significantly, $t(180) = -0.70$, Bonferroni-adjusted $p = 0.49$, $d = -0.10$.

***Downstream Consequences of LLM-Induced Shifts in Human Moral Inclinations***

We next examined whether LLM-induced shifts in participants' moral inclinations carried downstream consequences for their attitudes and decision-making on socio-political policies. To test this, we conducted two sets of mediation analyses using the PROCESS macro (Model 4; 10,000 bias-corrected bootstrapped resamples)[57], comparing (a) the D-LLM condition with the control condition and (b) the U-LLM condition with the control condition. In both analyses, we included both deontological and utilitarian inclinations as parallel mediators to test the indirect effects of D-LLM and U-LLM on participants' policy attitudes and voting intentions through changes in deontological and utilitarian inclinations. Below, we report on the focal results, with the full mediation analysis results provided in the Supplementary Information section I and Supplementary Tables 5 to 8.

Results show that D-LLM consistently increased participants' deontological inclination, which in turn led to systematically more *negative responses* to all socio-political policies. Mediation analyses showed that the indirect effects were significantly *negative* for 10 policy agreements and 11 voting intentions (Table 4). Specifically, increases in deontological inclination significantly *reduced* both policy agreement and voting intention for Abortion Freedom policy (indirect effect on policy agreement = -0.52, $SE = 0.17$, LLCI = -0.88, ULCI = -0.23; indirect effect on voting intention = -0.50, $SE = 0.18$, LLCI = -0.88, ULCI = -0.19), Prostitution Legalization and Regulation policy (indirect effect on policy agreement = -0.32, $SE = 0.12$, LLCI = -0.57, ULCI = -0.11; indirect effect on voting intention = -0.33, $SE = 0.12$, LLCI = -0.59, ULCI = -0.12), Euthanasia and Physician-Assisted Dying Legalization policy (indirect effect on policy agreement = -0.22, *SE* = 0.12, LLCI = -0.49, ULCI = -0.01; indirect effect on voting intention = -0.24, *SE* = 0.13, LLCI = -0.53, ULCI = -0.01), Compulsory Organ Donation policy (indirect effect on policy agreement = -0.43, *SE* = 0.15, LLCI = -0.75, ULCI = -0.17; indirect effect on voting intention = -0.36, *SE* = 0.15, LLCI = -0.71, ULCI = -0.10), Mandatory Vaccinations policy (indirect effect on policy agreement = -0.44, *SE* = 0.13, LLCI = -0.71, ULCI = -0.21; indirect effect on voting intention = -0.45, *SE* = 0.13, LLCI = -0.72, ULCI = -0.21), Government Surveillance policy (indirect effect on policy agreement = -0.29, *SE* = 0.13, LLCI = -0.58, ULCI = -0.07; indirect effect on voting intention = -0.22, *SE* = 0.11, LLCI = -0.45, ULCI = -0.0003), Firearm Ownership Restrictions policy (indirect effect on policy agreement = -0.22, *SE* = 0.12, LLCI = -0.45, ULCI = -0.003; indirect effect on voting intention = -0.26, *SE* = 0.12, LLCI = -0.49, ULCI = -0.03), Compulsory Military Service policy (indirect effect on policy agreement = -0.24, *SE* = 0.12, LLCI = -0.49, ULCI = -0.03; indirect effect on voting intention = -0.22, *SE* = 0.12, LLCI = -0.47, ULCI = -0.001), Comprehensive Terrorist Interrogation Methods policy (indirect effect on policy agreement = -0.31, *SE* = 0.12, LLCI = -0.60, ULCI = -0.10; indirect effect on voting intention = -0.27, *SE* = 0.12, LLCI = -0.53, ULCI = -0.07), and Quarantine Enforcement policy (indirect effect on policy agreement = -0.22, *SE* = 0.10, LLCI = -0.45, ULCI = -0.03; indirect effect on voting intention = -0.24, *SE* = 0.10, LLCI = -0.45, ULCI = -0.05). In addition, interacting with the D-LLM *reduced* voting intentions, but not policy agreements, for the Death Penalty policy (indirect effect = -0.30, *SE* = 0.12, LLCI = -0.57, ULCI = -0.08).

By contrast, although the U-LLM effectively strengthened participants' utilitarian inclination, this shift did not translate into significant indirect effects on agreement or voting intentions for any of the socio-political policies (Table 4).

Therefore, in line with our predictions, these findings indicate that although the U-LLM significantly increased participants' utilitarian inclination, this shift did not translate into systematic changes in their attitudes or decision-making. By contrast, the D-LLM exerted a strong and systematic effect on participants' responses to socio-political policies by elevating their deontological inclination, which led them to prioritize avoiding potential moral transgression and, consequently, reject the new policies.

**Discussion**

LLMs are now deeply embedded in everyday life across much of the world, influencing how humans think, decide, and behave across diverse domains. For this reason, extensive public and scholarly discussions have centered on which moral principles should guide machine behavior[1,2], with the aim of establishing an ethical framework for LLMs. However, our findings from two studies reveal a critical paradox: embedding moral principles in LLMs to guide their behavior may not only constrain machines but also shape human morality. By consistently generating responses aligned with their programmed principles, morally designed LLMs can influence users' moral inclinations, with downstream consequences for policy evaluations.

This work provides important and timely contributions to both theory and practice. First, it contributes to the growing body of research on LLMs and morality by empirically testing the competing theoretical perspectives regarding LLMs' influence on human morality. Specifically, our findings provide strong support for the perspective that morally programmed LLMs can systematically shift human moral inclinations, aligning them with the moral principles they reflect during interactive exchanges. Importantly, these effects were not merely transient. Rather than dissipating shortly after the interaction, LLMs' moral influence endured over a two-week period, suggesting that individuals internalized LLMs' arguments rather than merely complying superficially. Moreover, we also observed asymmetric temporal dynamics in LLMs' moral influence: while human deontological inclination tends to slightly regress toward baseline two weeks after interacting with morally programmed LLMs, their utilitarian inclination remains stable. These findings suggest that while morally programmed LLMs can meaningfully shift human moral inclinations, the durability of their effects depends on the type of moral inclination involved.

Second, our work contributes to the longstanding discussion surrounding AI ethics by uncovering a design paradox: embedding specific moral principles within LLMs may itself be ethically problematic. Given that morally programmed LLMs can influence users' moral inclinations through interactions, embedding predefined moral guidelines into these systems could raise significant concerns about covert *moral engineering*, referring to the subtle shaping of users' moral beliefs and judgments through system design. As a result, an LLM intended to be "ethical" may paradoxically become an instrument of moral engineering itself. We suggest that the discussion on ethical AI design should go beyond merely embedding explicit moral principles and grapple with the broader implications of how LLMs with encoded principles might influence human morality through everyday interactions. It is crucial to ask: Who decides which moral principles LLMs should follow? Should LLMs be programmed to promote certain moral principles, or should they adapt to individual users' ethical frameworks? These questions remain unresolved and warrant further interdisciplinary discussion among AI ethicists, policymakers, and researchers.

Third, our work contributes to the political psychology literature by identifying a novel pathway through which advanced technologies, i.e., LLM, can influence political attitudes and decision-making at scale. Prior research has primarily focused on how LLMs may affect political opinions through direct persuasive messaging about political issues[9,10], highlighting risks such as malicious prompting or adversarial manipulation. In contrast, the LLMs in our studies neither discussed political agendas nor attempted to persuade participants about specific policies. Instead, they engaged participants in discussions of moral dilemmas. Nonetheless, these interactions altered participants' underlying moral inclinations and, consequently, reshaped their judgments about socio-political policies. Thus, our findings show that LLMs may influence political attitudes not only through direct persuasion on political topics but also indirectly by shifting the moral lens through which individuals evaluate social and political questions. This suggests that morally programmed LLMs could have broad implications for public discourse and democratic decision-making even when they do not explicitly discuss political issues.

Notwithstanding, this research has some limitations that future work can fruitfully address. The first limitation, inherent to all moral dilemma research, is the closed-world assumption[58]. Moral dilemma paradigms typically assume that the described harm will unavoidably lead to the stated outcome and that no alternative actions are possible. However, participants may reject a particular action not due to deontological constraints but because they perceive the outcome as uncertain or unrealistic. Nonetheless, past research has shown that judgments in moral dilemmas can reflect real-world decision-making[59] and capture broader moral concerns beyond the specific scenarios presented[47]. Therefore, despite this limitation, moral dilemmas remain an essential tool for advancing research in moral psychology.

Furthermore, while our work demonstrates the main effect of LLMs' moral influence on human morality, it leaves open the question of whether this effect varies based on individual differences. Research on morality has long demonstrated that personality traits, cognitive styles, and ideological orientations shape moral decision-making and susceptibility to persuasion[60,61].

Thus, it is likely that LLMs' influence on moral inclinations may vary across individuals, depending on their characteristics. Although our studies rely on random assignment to conditions, which reduces concerns that such individual differences systematically bias our results, we did not examine how these factors may moderate LLM influence. Investigating how individual differences affect the extent to which LLMs influence human moral inclinations represents a promising direction for future research.

## Methods

All studies were approved by the Ethics Review Group at the corresponding author's institution. The design, sample size, and analysis plans for all studies were pre-registered before data collection, with Study 1 pre-registered on October 13, 2024 (https://aspredicted.org/h3hq-zkfq.pdf), and Study 2 pre-registered on October 31, 2024 (https://aspredicted.org/h5sd-grf6.pdf). Study materials and pre-registrations for all studies are also available on the Open Science Framework: https://osf.io/emrjb/overview?view_only=18a399dc2b03452daf61633e3092c4b6. The only deviation from the pre-registered analysis plans occurred in Study 2. We mistakenly pre-registered multilevel mediation analyses. However, the data collected were single-level rather than multilevel, and thus we conducted mediation analyses instead. All participants provided informed consent and were compensated at a standard rate of £6 per hour.

### *Process Dissociation Method*

Across all studies, we employed the PD method to measure human moral inclinations. The PD method estimates moral inclinations using a moral dilemma battery comprising 10 incongruent and 10 congruent dilemmas spanning a wide range of topics (see Supplementary Table 1 for dilemma descriptions). Specifically, incongruent dilemmas present a conflict between deontological and utilitarian inclinations. In these dilemmas, the deontological perspective emphasizes the rejection of harm, while the utilitarian perspective focuses on maximizing overall outcomes. This creates a tension between the two responses. Congruent dilemmas, in contrast, describe similar harmful actions that yield less advantageous outcomes, resulting in both deontological and utilitarian approaches agreeing on the need to reject the harm. By comparing responses across these dilemma types, the PD method yields separate estimates for participants' deontological and utilitarian inclinations.

*Study 1*

We collected data from 160 participants using a three-wave design on Prolific, aiming for 80 participants per condition. Of these, 34 participants (attrition rate: 21.25%) failed to complete all three waves and were excluded. Following prior work using the PD method[27], we further excluded one participant who had 100% utilitarian responses at Time 3 due to the mathematical constraints of the PD method (see Supplementary Information section B for details). Additionally, one participant was excluded for failing to interact with the assigned LLM. The final sample consisted of 124 participants (68 males, 55 females, 1 non-binary/third gender; $M_{age}$ = 32.17, $SD_{age}$ = 9.93). A sensitivity analysis using G*Power[62] indicated that, with our final sample ($N = 124$), we had 80% power ($\alpha = 0.05$) to detect a small to medium within-subject effect of Cohen's $f = 0.18$, and a medium between-subject effect of Cohen's $d = 0.45$.

Data were collected at three time points to establish baseline assessment at Time 1, immediate effects at Time 2 (one week after Time 1), and enduring effects at Time 3 (three weeks after Time 1). At Time 1, participants independently completed the PD method's moral dilemma battery in a randomized order, followed by demographic questions (i.e., age and gender). At Time 2 (one week after Time 1), participants were randomly assigned to one of two

LLM conditions: in the D-LLM condition, participants interacted with the D-LLM, while in the U-LLM condition, they interacted with the U-LLM. Aside from the LLM itself, all participants followed the same procedure. They engaged in discussions with the assigned LLM on 20 moral dilemmas. Each dilemma was presented on a separate survey page with the validated LLM integrated into the interface. For each dilemma, participants first read the description and interacted with the assigned LLM, and then made a final judgment about whether the described behavior was appropriate. The order of the dilemmas was randomized for each participant. At Time 3 (two weeks after Time 2), participants were re-invited to complete the PD method's moral dilemma battery independently, with dilemmas presented in randomized order. Means, standard deviations, and correlations for Study 1 are available in Supplementary Table 3.

*Study 2*

We recruited 300 participants from the United States via Prolific, aiming for 100 participants per condition. We chose to include only participants from the United States with English as their first language for two reasons: (a) to ensure fluent communication and high-quality interactions with the LLMs, and (b) because several policy topics examined in the study, such as gun control, are specifically embedded within the U.S. socio-political context. Twenty-two participants who failed the attention check and four participants who produced 100% utilitarian responses were removed (see Supplementary Information section B for details), leaving a final sample of 274 participants (128 males, 143 females, one non-binary/third gender, and two participants chose not to disclose their gender; $M_{age}$ = 40.12, $SD_{age}$ = 12.93). A sensitivity analysis for ANOVA using G*Power [62] indicated that with our final sample, we could detect a small to medium between-subject effect of Cohen's $f$ = 0.19 with 80% power ($\alpha$ = 0.05).

Participants were randomly assigned to one of three conditions: D-LLM, U-LLM, or a control condition. All participants first completed the PD method's moral dilemma battery, presented in a randomized order. Consistent with Study 1, in the two LLM conditions, each dilemma was presented on a separate survey page with the assigned LLM integrated. Participants in the D-LLM condition interacted with the D-LLM, while those in the U-LLM condition interacted with the U-LLM. For each dilemma, participants read the description, engaged in real-time interaction with their assigned LLM, and made the final judgment. In the control condition, participants responded to the PD method's moral dilemma battery independently, without any exposure to LLMs. Participants' deontological and utilitarian inclinations were assessed using the PD method.

To examine the downstream consequences of LLM-induced changes in human moral inclinations, participants in all conditions then completed the socio-political policy task. They were asked to imagine that their local government was considering enacting a set of public policies. After reviewing each of the 11 policy proposals, they rated their level of agreement and indicated their likelihood of voting in favor. We developed these 11 policy proposals based on contemporary social issues that are frequently discussed in major news outlets (Table 3). Our goal was to capture a diverse range of policy domains, including public health, civil liberties, public security, and criminal justice. This broad scope allowed us to investigate whether LLM-induced shifts in human moral inclinations influence individuals' attitudes and decision-making across various policy areas.

To measure participants' policy attitudes and decisions, participants were asked to rate their agreement with each policy proposal (policy agreement) on a 7-point Likert scale ranging from "1" (strongly disagree) to "7" (strongly agree) and indicated their willingness to vote in favor of each policy (voting intention) on a 7-point Likert scale ranging from "1" (I will definitely not vote in favor of this policy) to "7" (I will definitely vote in favor). By capturing both attitudinal agreement and behavioral intentions, this approach provides a comprehensive understanding of whether LLM-induced changes in human moral inclinations influence not only participants' policy attitudes but also their real-world decision-making tendencies. Means, standard deviations, and correlations for Study 2 are available in Supplementary Table 4.

**Data Availability**

The data for all studies are available on the Open Science Framework:

https://osf.io/emrjb/overview?view_only=18a399dc2b03452daf61633e3092c4b6.

**Code Availability**

The analysis code for all studies is available on the Open Science Framework:

https://osf.io/emrjb/overview?view_only=18a399dc2b03452daf61633e3092c4b6.

**Competing interests**

**Supplementary Information**

This manuscript contains supplementary information.

**Fig. 1. Overview of Studies.**

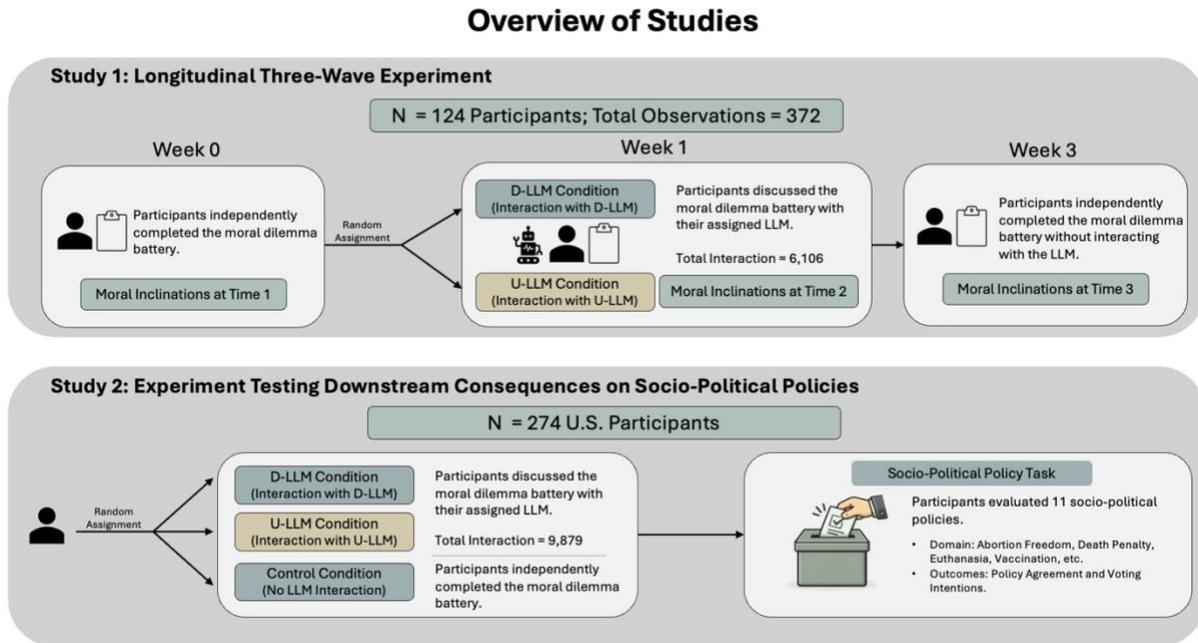

**Fig. 2. Changes in Participants' Moral Inclinations from Time 1 to Time 3 in Study 1.**

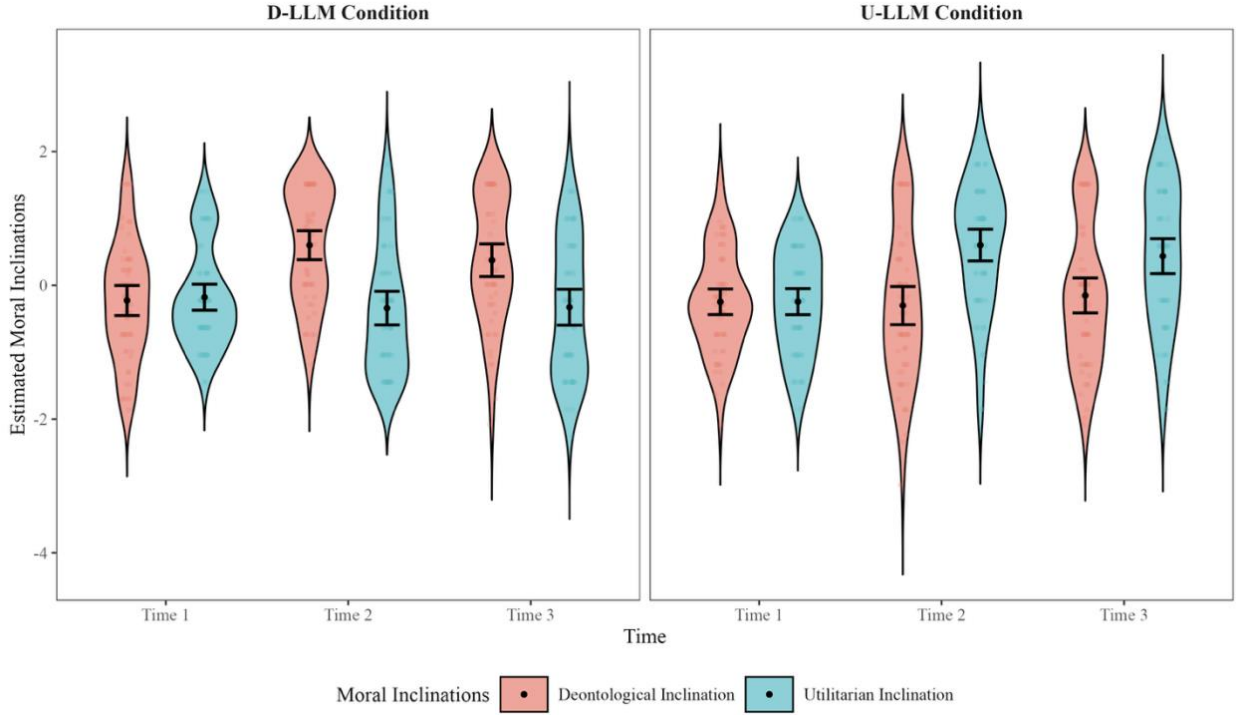

*Note.* Error bars represent 95% confidence interval.

**Fig. 3. Participants' Moral Inclinations in Each of the Three Conditions in Study 2.**

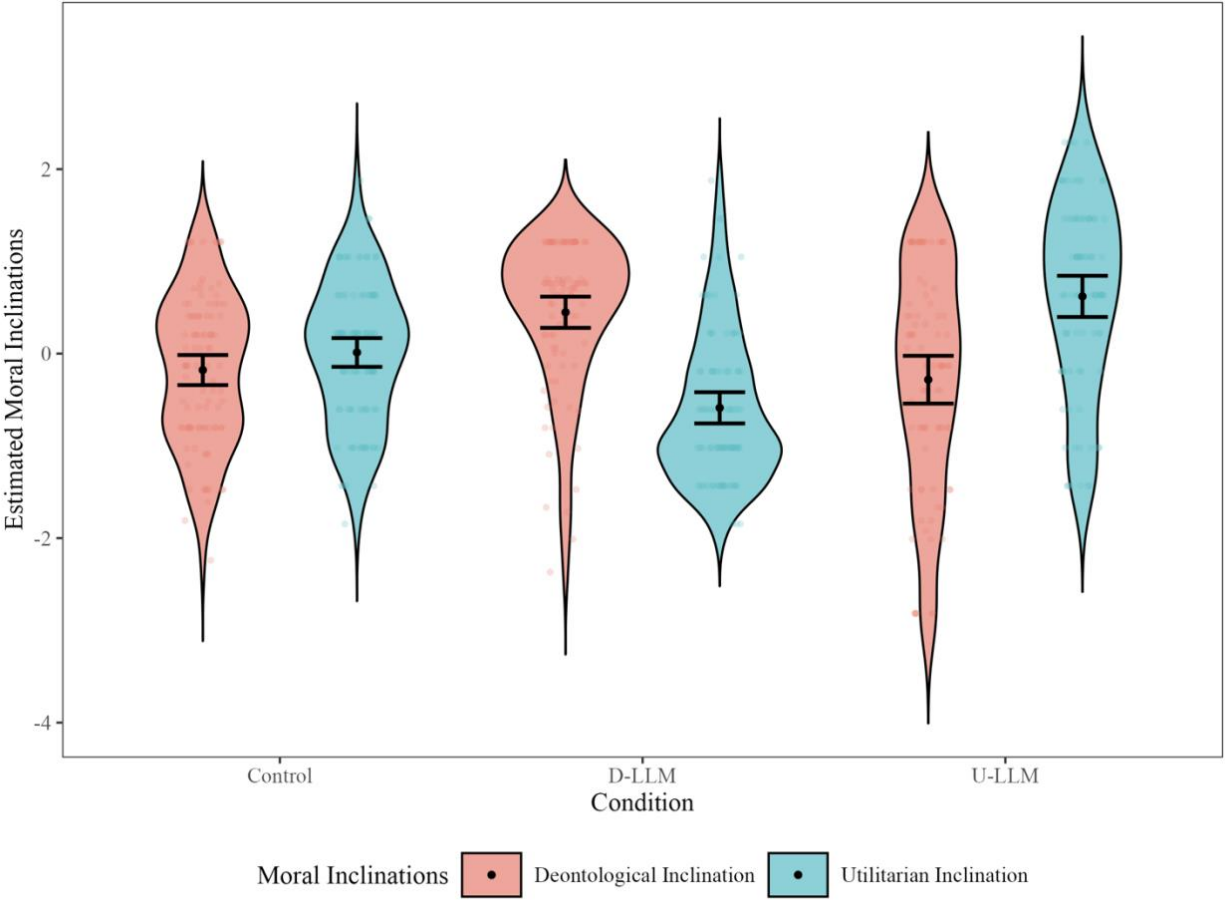

*Note.* Error bars represent 95% confidence interval.

**Table 1. Descriptive statistics for Study 1.**

|  |  | Time 1 | | Time 2 | | Time 3 | |
| --- | --- | --- | --- | --- | --- | --- | --- |
| Condition | Moral Inclinations | M | SD | M | SD | M | SD |
| D-LLM Condition | Deontological Inclination | -0.23 | 0.87 | 0.60 | 0.84 | 0.37 | 0.94 |
|  | Utilitarian Inclination | -0.18 | 0.75 | -0.34 | 0.97 | -0.33 | 1.04 |
| U-LLM Condition | Deontological Inclination | -0.25 | 0.77 | -0.30 | 1.14 | -0.15 | 1.05 |
|  | Utilitarian Inclination | -0.24 | 0.78 | 0.60 | 0.95 | 0.44 | 1.05 |

*Note. M* and *SD* are used to represent mean and standard deviation, respectively.

**Table 2. Descriptive statistics for Study 2.**

| Condition | Moral Inclinations | M | SD |
|---|---|---|---|
| D-LLM Condition | Deontological Inclination | 0.45 | 0.82 |
|  | Utilitarian Inclination | -0.59 | 0.82 |
| U-LLM Condition | Deontological Inclination | -0.28 | 1.20 |
|  | Utilitarian Inclination | 0.62 | 1.04 |
| Control Condition | Deontological Inclination | -0.18 | 0.81 |
|  | Utilitarian Inclination | 0.01 | 0.77 |

*Note. M* and *SD* are used to represent mean and standard deviation, respectively.

**Table 3. Socio-Political Policies in Study 2.**

| Socio-Political Policy | Description |
|---|---|
| Abortion Freedom | This new policy will ensure that pregnant individuals have full autonomy and freedom of choice regarding abortion. |
| Mandatory Vaccinations | In this new policy, all individuals are required to receive vaccinations for preventable infectious diseases as determined necessary by public health authorities. Exemptions are granted exclusively to those with valid medical reasons substantiated by certified healthcare professionals. |
| Quarantine Enforcement | This new policy will mandate quarantine measures for individuals living in areas affected by infectious disease outbreaks. Infected individuals or those suspected of infection are required to remain in designated quarantine facilities or their residences for a specified period. |
| Compulsory Organ Donation | In this new policy, all eligible individuals are required to donate their organs upon death, except in cases where the organs are deemed medically unusable. |
| Euthanasia and Physician-Assisted Dying Legalization | This new policy will legalize euthanasia and physician-assisted dying for competent adults who voluntarily request it. |
| Comprehensive Terrorist Interrogation Methods | This new policy will allow the use of all available methods, including torture, in the interrogation of terrorists. |
| Death Penalty | This new policy will authorize judges to impose the death penalty as the ultimate form of punishment for individuals convicted of the most severe crimes under clearly defined and stringent legal conditions. |

| Firearm Ownership Restrictions | This new policy will prohibit the personal ownership of firearms, except for law enforcement officers and other authorized individuals. |
|---|---|
| Compulsory Military Service | In this new policy, all eligible individuals are required to undertake compulsory military service when mandated by the government. Exceptions are granted solely to those who are determined to be physically or mentally unfit for service based on established medical and psychological evaluations. |
| Government Surveillance | This new policy will authorize government authorities to monitor security-threatening communications and activities by collecting data from emails, phone calls, internet usage, and surveillance in public spaces through facial recognition and license plate recognition systems. |
| Prostitution Legalization and Regulation | This new policy will legalize and regulate prostitution, encompassing licensing requirements, operational standards, and oversight mechanisms. |

**Table 4. Results of Mediation Analyses in Study 2.**

|  | Indirect Effect of D-LLM via Deontological Inclination | | Indirect Effect of U-LLM via Utilitarian Inclination | |
|---|---|---|---|---|
| **Socio-Political Policy** | **Policy Agreement** | **Voting Intention** | **Policy Agreement** | **Voting Intention** |
| Abortion Freedom | **-0.52** [-0.88, -0.23] | **-0.50** [-0.88, -0.19] | 0.07 [-0.17, 0.29] | 0.08 [-0.15, 0.30] |
| Mandatory Vaccinations | **-0.44** [-0.71, -0.21] | **-0.45** [-0.72, -0.21] | 0.03 [-0.20, 0.25] | 0.02 [-0.21, 0.26] |
| Quarantine Enforcement | **-0.22** [-0.45, -0.03] | **-0.24** [-0.45, -0.05] | 0.12 [-0.05, 0.30] | 0.13 [-0.04, 0.32] |
| Compulsory Organ Donation | **-0.43** [-0.75, -0.17] | **-0.36** [-0.71, -0.10] | 0.02 [-0.18, 0.24] | -0.005 [-0.20, 0.22] |
| Euthanasia and Physician-Assisted Dying Legalization | **-0.22** [-0.49, -0.01] | **-0.24** [-0.53, -0.01] | 0.10 [-0.10, 0.33] | 0.09 [-0.11, 0.31] |
| Comprehensive Terrorist Interrogation Methods | **-0.31** [-0.60, -0.10] | **-0.27** [-0.53, -0.07] | 0.07 [-0.14, 0.29] | 0.09 [-0.11, 0.31] |
| Death Penalty | -0.20 [-0.51, 0.03] | **-0.30** [-0.57, -0.08] | 0.12 [-0.10, 0.38] | 0.11 [-0.09, 0.36] |
| Firearm Ownership Restrictions | **-0.22** [-0.45, -0.0003] | **-0.26** [-0.49, -0.03] | -0.14 [-0.42, 0.07] | -0.13 [-0.40, 0.09] |
| Compulsory Military Service | **-0.24** [-0.49, -0.03] | **-0.22** [-0.47, -0.001] | -0.08 [-0.27, 0.11] | -0.04 [-0.23, 0.17] |
| Government Surveillance | **-0.29** [-0.58, -0.07] | **-0.23** [-0.51, -0.03] | -0.03 [-0.22, 0.17] | 0.004 [-0.18, 0.18] |
| Prostitution Legalization and Regulation | **-0.32** [-0.57, -0.11] | **-0.33** [-0.59, -0.12] | 0.07 [-0.13, 0.29] | -0.01 [-0.22, 0.20] |

Note. Bolded values indicate statistically significant effects ($p < 0.05$). 95% confidence intervals are presented in brackets below each effect estimate.